# Moiré-induced electronic structure modifications in monolayer $V_2S_3$ on Au(111)

Umut Kamber[1], Sahar Pakdel[2], Raluca-Maria Stan[2], Anand Kamlapure[1], Brian Kiraly[1], Fabian Arnold[2], Andreas Eich[1], Arlette S. Ngankeu[2], Marco Bianchi[2], Jill A. Miwa[2], Charlotte E. Sanders[3], Nicola Lanatà[2], Philip Hofmann[2], Alexander A. Khajetoorians[1]*

1. Institute for Molecules and Materials, Radboud University, Nijmegen 6525AJ, The Netherlands

2. Department of Physics and Astronomy, Interdisciplinary Nanoscience Center, Aarhus University, 8000 Aarhus C, Denmark

3. Central Laser Facility, STFC Rutherford Appleton Laboratory, Harwell, Didcot OX11 0QX, United Kingdom

ABSTRACT: There is immense interest in how the local environment influences the electronic structure of materials at the single layer limit. We characterize moiré induced spatial variations in the electronic structure of in-situ grown monolayer $V_2S_3$ on Au(111) by means of low temperature scanning tunneling microscopy and spectroscopy. We observe a long-range modulation of the integrated local density of states (LDOS), and quantify this modulation with respect to the moiré superstructure for multiple orientations of the monolayer with respect to the substrate. Scanning tunneling spectroscopy reveals a prominent peak in the LDOS, which is shifted in energy at different points of the moiré superstructure. Comparing *ab initio* calculations with angle-resolved photoemission, we are able to attribute this peak to bands that exhibit a large out-of-plane *d*-orbital character. This suggests that the moiré driven variations in the measured density of states is driven by a periodic modulation of the monolayer-substrate hybridization.

INTRODUCTION

Moiré superstructures, constructed from monolayers of van der Waals (vdW) materials like graphene, have emerged as a tunable platform to artificially create novel quantum states of matter[1-5]. As demonstrated for structures involving monolayer graphene, moiré interference can lead to long-range structural and electronic modulations, which can be imaged with scanning tunneling microscopy/spectroscopy (STM/STS) and low energy electron diffraction (LEED)[6, 7]. Such moiré superstructures, for example involving monolayer graphene, lead to various modifications of the band structure[8], such as the minigap and replica band features observed in angle-resolved photoemission spectroscopy (ARPES)[9]. The moiré superstructure and its associated structural/electronic modifications can be quenched by decoupling the monolayer via intercalation, leading to quasi-free-standing electronic structure[10, 11]. As the family of tunable moiré materials grows[12-15], it is important to understand how moiré superstructures can modify the structural, electronic, and magnetic properties of vdW monolayers at the atomic length scale.

One class of vdW materials, transition metal dichalcogenides (TMDs), have been synthesized at the monolayer limit on various surfaces leading to moiré superstructures[16-22]. In the case of TMDs grown on metallic surfaces, strong interactions with the substrate, typically resulting from hybridization and modified charge screening, lead to pronounced modifications in the electronic structure[23, 24]. For the $MoS_2$ monolayer on Au(111), these strong interactions lead to a renormalization of the band gap[22, 25, 26] when compared to the quasi-free-standing case[27]. Similarly, for the $TaS_2$ monolayer grown on Au(111), interactions with the underlying substrate suppress the expected charge density wave (CDW) phase[16, 23, 28], which can remain robust in $TaS_2$ monolayers prepared on other substrates[29]. Yet, for many TMD monolayers where the moiré superstructure shows strong signatures in STM and LEED, the moiré superstructure minimally perturbs the band structure observed in ARPES[16, 17]. Because many TMD monolayers show non-trivial electronic phases[29-31], it is important to study how long-range structural and electronic modifications induced by moiré superstructures modify the electronic structure.

Here, we study the structural and electronic properties of monolayer $V_2S_3$ on Au(111) to quantify the interplay between the moiré superstructure, the geometrical structure, and the electronic structure. Utilizing STM/STS, we observe well-defined, long-range modulations in constant-current imaging, i.e. the integrated density of states, reminiscent of a CDW. By investigating monolayer domains with varying rotation angle with respect to the Au(111) substrate, we confirm the persistence of these real-space modulations at all observed angles. Using detailed moiré simulations, we capture these modulations by considering higher-order wave vectors from the atomic lattice. Concomitantly, we use STS to identify a strong peak-like feature in the LDOS near $E_F$, which shifts in energy depending on the location within the moiré unit cell. In order to identify the peak seen in STS, we computed the band structure within density functional theory (DFT). By calibrating the effective doping using ARPES, we link the peak in the LDOS with the onset of weakly dispersing bands arising from the intrinsic $V_2S_3$ monolayer, with a substantial out-of-plane orbital character. We suggest that the out-of-plane character of the $V_2S_3$ bands enables them to couple to the possible structural modulations of the moiré superstructure, inducing the observed spatial variations in the electronic structure.

RESULTS AND DISCUSSION

Characterization of the atomic structure of the $V_2S_3$ monolayer is shown in Figure 1 using high-resolution constant-current imaging. A two-fold symmetric structure of the top-layer S atoms is visible; within this layer, the S atoms form a rectangular unit cell (***a,b***) with two S (red and gray atoms in Figure 1d) atoms along ***b***, while the bottom S layer (pink atoms in Figure 1d) maintains a configuration similar to that of $VS_2$ . The atomically resolved constant-current image in Figure 1c reveals a unit cell with lattice constants $|\boldsymbol{a}|$ = 0.31±0.03 nm and $|\boldsymbol{b}|$ = 0.87±0.03 nm. The unit cell of the $V_2S_3$ is indicated with a dashed green rectangle in Figure 1c, which is consistent with the model in Figure 1d, determined using room-temperature STM and X-ray photoelectron diffraction[32].

Concurrently with the atomic structure, additional longer-range modulations can be seen in constant-current imaging. Examining a fast Fourier transform (FFT) of the image in Figure 1a (Figure 1b), we observe many well-defined peaks corresponding to long-range features observed in real space (the reciprocal space points of the atomic lattice are highlighted by green circles). In constant-current images, a weaker moiré superstructure was observed compared to TMD layers grown on Au(111)[16, 22]. We label the first-order features in reciprocal space, namely, the largest real-space periodicities that are responsible for the observed moiré superstructure, A and B. In addition to these features, we also observe periodic modulations in constant-current imaging with characteristic real-space length scales longer than the atomic lattice, yet smaller than the moiré lattice. These multiple higher-order features are also visible in the FFT, two of which we label C and D (comprised of linear combinations of A, and B). While higher order features related to the moiré superstructure are often visible in LEED for families of sulfide-based TMDs grown on Au(111), there are commonly no strong indications of such higher order features in STM measurements in contrast to what is observed here[16]. Yet for $V_2S_3$ on Au(111), LEED measurements were not able to identify a moiré superstructure or any related higher-order features[32].

To further study these higher order features, we characterized different orientations of the $V_2S_3$ layer with respect to the substrate using atomic resolution imaging. In Figure 2a, the stacking angle between $\textbf{\textit{a}}$ and the <110> directions of Au(111) is $\theta = 0°$, measured by comparing the atomic lattice of $V_2S_3$ with the herringbone reconstruction of underlying Au(111). Two other monolayer islands with measured rotation angles of $\theta = 1.3° \pm 0.5°$ and $\theta = 4.2° \pm 1.0°$ are shown in Figure 2b and Figure 2c. However, approximately 80% of the observed $V_2S_3$ islands were aligned with the stacking angle of $\theta = 0°$ (forming three rotational domains at 120° with respect to each other consistent with the three-fold symmetry of the Au(111) substrate), indicating that there is a preferred growth orientation for the $V_2S_3$ monolayer on Au(111) (see Supporting Information Figure S2). As seen in the FFTs in Figure 2d-f, both moiré superstructures and higher-order features are also present in the misaligned cases.

To understand the origin of the observed higher-order features and how these features relate to the geometrical influence of the moiré superstructure, we simulated the expected moiré superstructures by representing the top S layer of $V_2S_3$ and the topmost Au(111) layer as sinusoidal functions, and multiplying them using a rotation angle defined as the angle between $\textbf{\textit{a}}$ and <110>[33]. We used higher-order sinusoidal functions for each lattice to visualize all the possible higher-order moiré modes (see Supporting Information section 1 for details). The resulting FFT of the simulations and their corresponding real space data are shown in Figure 2g-i for each observed moiré superstructure. Our simulations successfully reproduce all of the reciprocal space features for the $\theta = 0°$ moiré superstructure, as well as nearly all the features in the rotated cases when we include higher-order functions for both $V_2S_3$ and Au(111) (Figure 2g). Thus, we can confirm that the longer-range features observed in constant-current imaging stem from structural variations in the film.

To study local modifications in the electronic structure, we performed STS at different positions within the moiré superstructure (Figure 3). A signature of the $V_2S_3$ monolayer is a distinct peak near $E_F$. We observe that the position of this peak strongly varies depending on the spatial location within the moiré superstructure, as seen in d$I$/d$V$ spectra taken on the apparent moiré minima (light blue) and apparent moiré maxima (dark blue) in Figure 3 (see also Supporting Information Figure S4). This can also be seen in spatially dependent maps of the differential conductance near the Fermi energy, leading to complex patterns (see Supporting Information Figure S3). By contrast, peak-like features seen in the LDOS for $TaS_2$/Au(111) do not show significant shifts in energy within the moiré superstructure[23]. On the other hand, moiré induced modifications to the LDOS were observed in monolayer $MoS_2$ on Au(111), in particular energy ranges, which the authors attributed to possible differences in hybridization or electronic screening[34]. We suggest that the shifts in energy of the LDOS peak-like feature (Figure 3) may result from different hybridization strengths associated with the orbital character of the bands in this energy range; as shown in Figure S5 , the orbital character contains an out-of-plane component that may be sensitive to the variations in interlayer distance between the monolayer and the substrate.

To elucidate the electronic properties of the system, including the origin of the observed peak in the LDOS, we performed DFT calculations of the free-standing monolayer $V_2S_3$. In Figure 4a, we show both the calculated band structure along the high-symmetry path indicated in the inset and the total DOS (Figure 4b) of the free-standing monolayer $V_2S_3$. In order to account for the doping induced by the substrate, we shifted the calculated Fermi energy by assuming a rigid offset. By comparing the ARPES spectra with the monolayer DFT bands (Fig. 4a-c) at the $\Gamma$ point and the Fermi contour (Fig. 4d), we observed that a satisfactory agreement is obtained by shifting the DFT bands by about 180 meV-220 meV. We note that in Figure 4 the theoretical Fermi level $E_F$ (DFT) is lowered by 180 meV, and we refer to such shifted Fermi level as $E_F$ (Exp). The comparison between the experimental and theoretical constant energy contours (including three 120° rotations corresponding to the three experimental domain rotations) with this offset Fermi level ($E_F$ (Exp)) is shown in Figure 4d (see also Supporting Information Figure S6). The photoemission process is highly dependent on matrix elements; thus, not all features predicted by theory are visible in the ARPES data. However, the main qualitative features of the measured energy contours (EC) are captured by the DFT calculations. Further details of EC are discussed in the supplemental material.

Comparing the calculated total DOS (Figure 4b) to the observed peak in STS (Figure 3), we find good agreement. The calculated DOS of the native monolayer in Figure 4b contains sharp features at the shifted Fermi level, similar to features that are observed in d$I$/d$V$ measurements in Figure 3. To understand the orbital texture of the bands contributing to the calculated DOS features at $E_F$ (Exp), we projected the total DOS onto the vanadium $d$-orbitals (see Supporting Information Figure S5). From this decomposition, it emerges that these peaks in the total DOS have a substantial contribution from orbitals with out-of-plane character. We suggest that such out-of-plane orbital character may make the bands more susceptible to influence from the physical corrugation of the monolayer due to the moiré superstructure. The moiré superstructure can modify the hybridization of the monolayer bands with the substrate by varying the interlayer distance, resulting in the observed shift in the peak position in Figure 3. As this behavior was also observed on misaligned moiré superstructures (Figure 3b), it appears to be related to the physical corrugation of the monolayer.

Both CDWs and magnetism have been observed in other V-based TMDs[35-39] and predicted in $VS_2$[40-42]. At temperatures down to 1.3 K, the enhanced DOS peak near $E_F$ is consistently observed, in contrast with common expectations of a minimum in the DOS near the Fermi level in CDW systems. In order to investigate the interplay between the electronic and orbital degrees of freedom in our moiré system, we calculated the nesting function and the electronic susceptibility of the free-standing monolayer $V_2S_3$ (see Supporting Information section 5). Interestingly, both of these functions display a peak in the vicinity of one of the second-order reciprocal-space points of the moiré superstructure (indicated as the C-point in Figure 1, and Figure S7). This suggests the possibility of a connection between the formation of the moiré superstructure and the underlying electronic instabilities of the free-standing layer in this system. However, from this calculation alone, we cannot draw any direct conclusion concerning this point. Likewise, in order to draw any experimental conclusions about the tendency toward CDW formation, further work involving temperature-dependent STM measurements is required.

CONCLUSIONS

In conclusion, we studied the electronic structure of monolayer $V_2S_3$ grown on Au(111). Constant-current imaging revealed long-range modulations persistent at various rotation angles; as these modulations were captured in moiré simulations, they were attributed to higher-order modes from the moiré superstructure. Tunneling spectroscopy revealed a strong peak feature in the LDOS near $E_F$, whose energy varied depending on the position within the moiré superstructure. Using ARPES, in conjunction with DFT calculations, we were able to relate the observed peak in STS to $V_2S_3$ bands with hybridization effects mainly due to the out-of-plane vanadium $d$-orbitals. We suggest that the moiré induced modulation of this peak could be due to coupling between the out-of-plane band character and the physical moiré corrugation. Future measurements on the temperature dependent evolution of the system could provide valuable insight into the interplay between the moiré modulated electronic structure and possible electronic instabilities in this material.

METHODS

**Sample Preparation.** $V_2S_3$ samples were prepared in-situ by the method described previously[32] . A clean Au(111) surface was first prepared by sputtering $Ar^+$ (1.5 kV) ions and subsequent annealing at 600 ºC. Vanadium was then deposited by e-beam evaporation onto the substrate, which was held at room temperature. Following the deposition, the sample was annealed to $T = 450$ ºC in an atmosphere of $p = 5$ x $10^{-5}$ mbar $H_2S$ partial pressure, yielding monolayer $VS_2$. $V_2S_3$ was then obtained by annealing the $VS_2$ to $T = 600$ ºC at p < $1x10^{-9}$ mbar. The samples were in-situ transferred into the STM without breaking vacuum.

**STM/STS measurements.** All STM/STS measurements were performed under ultrahigh vacuum conditions with home-built low-temperature setup, operating at a base temperature of $T = 1.3$ K and 4.2 K, with the capability of applying a magnetic field perpendicular to the surface up to B = 9 T[43]. All STM images in the main text and the Supporting Information were acquired by using constant-current feedback with the bias applied to the sample ($V_s$). Electrochemically etched tungsten (W) tips were used after in-situ treatment by electron bombardment. d$I$/d$V$ measurements were performed using a lock-in technique with a modulation bias ($V_{mod}$) at a frequency ($f_{mod}$) of 880 Hz added to the bias signal. Fast Fourier transforms of the real space STM images were calculated using MATLAB software. All the FFT images shown in this study have minimum 0.41 $nm^{-1}$/pixel resolution.

**ARPES measurements.** All ARPES measurements were acquired using a photon energy of 94 eV at the SGM3 endstation of the ASTRID2 synchrotron light source in Aarhus, Denmark. The energy and angular resolution were better than 40 meV and 0.2°, respectively [44].

**DFT calculations.** The DFT calculations were performed using the Vienna *ab initio* Simulation Package (VASP) code[45, 46]. The exchange–correlation potentials were described through the Perdew–Burke–Ernzerhof (PBE) functional within the generalized gradient approximation (GGA) formalism[47]. A plane wave basis set was used with a cutoff energy of 400 eV on a 7 x 21 Monkhorst-Pack k-point mesh[48]. Lattice constants reported by Arnold *et.al*[32] was used as the reference and was modified less than 10% to achieve a commensurate moiré lattice with the Au(111) substrate. A vacuum region of 25 Å along the z direction was used in order to minimize the interaction between the periodic repetitions of the cell.

ASSOCIATED CONTENT

**Supporting Information**

Details of the moiré simulations, additional real space characterization by STM and STS, DFT calculations for projected DOS of the *d*-orbitals of vanadium, additional ARPES results, and calculated electronic susceptibility of the free-standing monolayer $V_2S_3$.

AUTHOR INFORMATION

**Corresponding Author**

Alexander A. Khajetoorians - Institute for Molecules and Materials, Radboud University, Nijmegen 6525AJ, The Netherlands; Email: a.khajetoorians@science.ru.nl

**Author Contributions**

The manuscript was written through contributions of all authors. All authors have given approval to the final version of the manuscript.

ACKNOWLEDGMENT

B.K. acknowledges NWO-VENI project "Controlling magnetism of single atoms on black phosphorus" with project number 016.Veni.192.168. S.P. acknowledges support from Spanish MINECO for the computational resources provided through Grant FIS2016-80434-P. A.K. and A. A. K. acknowledges the VIDI project: "Manipulating the interplay between superconductivity and chiral magnetism at the single atom level" with Project No. 680-47-534, which is financed by NWO. U.K., A.E., and A.A.K. acknowledge funding from NWO. A. A. K acknowledges support from the European Research Council (ERC) under the European Union's Horizon 2020 research and innovation programme (Grant Agreement No. 818399, SPINAPSE). This work was supported by VILLUM FONDEN via the Centre of Excellence for Dirac Materials (Grant No. 11744). J.A.M. acknowledges financial support from the Danish Council for Independent Research, Natural Sciences under the Sapere Aude program (grant no. DFF-6108-00409), and from Aarhus University Research Foundation.

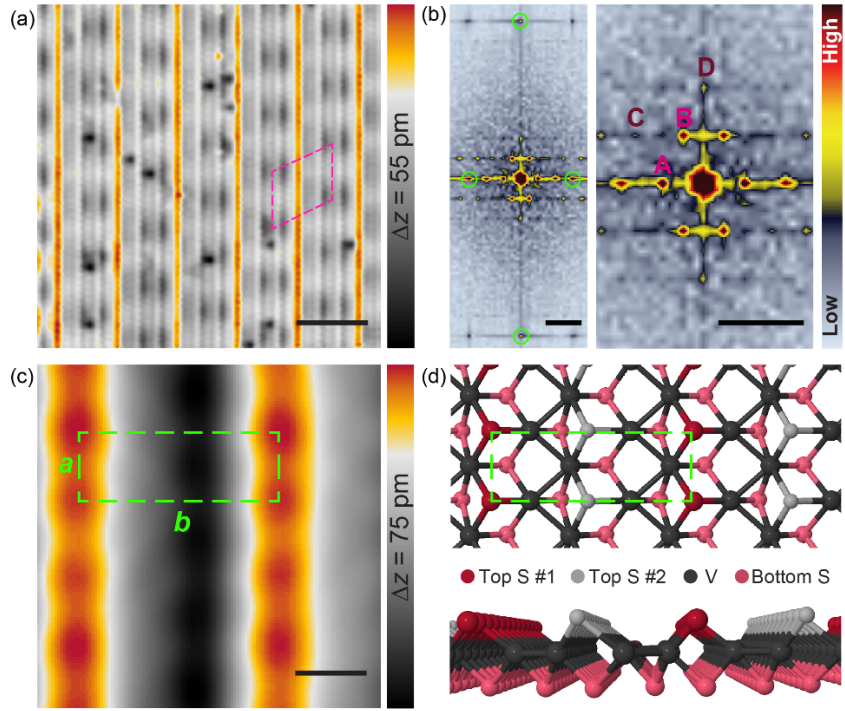

Figure 1. (a) Large scale constant-current STM image of the surface, illustrating the two-fold symmetric atomic lattice of top S atoms, as well as a moiré superstructure caused by the lattice mismatch between $V_2S_3$ monolayer and Au(111) substrate ($V_s$ = 1 V, $I_t$ = 40 nA, $T$ = 1.3 K, scale bar = 3 nm). (b) Corresponding FFT of the image in (a) (scale bar = 5 nm$^{-1}$), with a close-up view of the center (scale bar = 5 nm$^{-1}$). Green circles in the FFT images highlight the reciprocal lattice points of the atomic lattice; the corresponding real-space unit is drawn in (c). The real space unit cell of the moiré is plotted with dashed magenta lines in (a). (c) Atomically resolved constant-current image of ML $V_2S_3$ revealing the atomic unit cell ($V_s$ = 3 mV, $I_t$ = 1 nA, $T$ = 1.3 K, scale bar = 300 pm). (d) Structural model of ML $V_2S_3$/Au(111) (adapted from [32]).

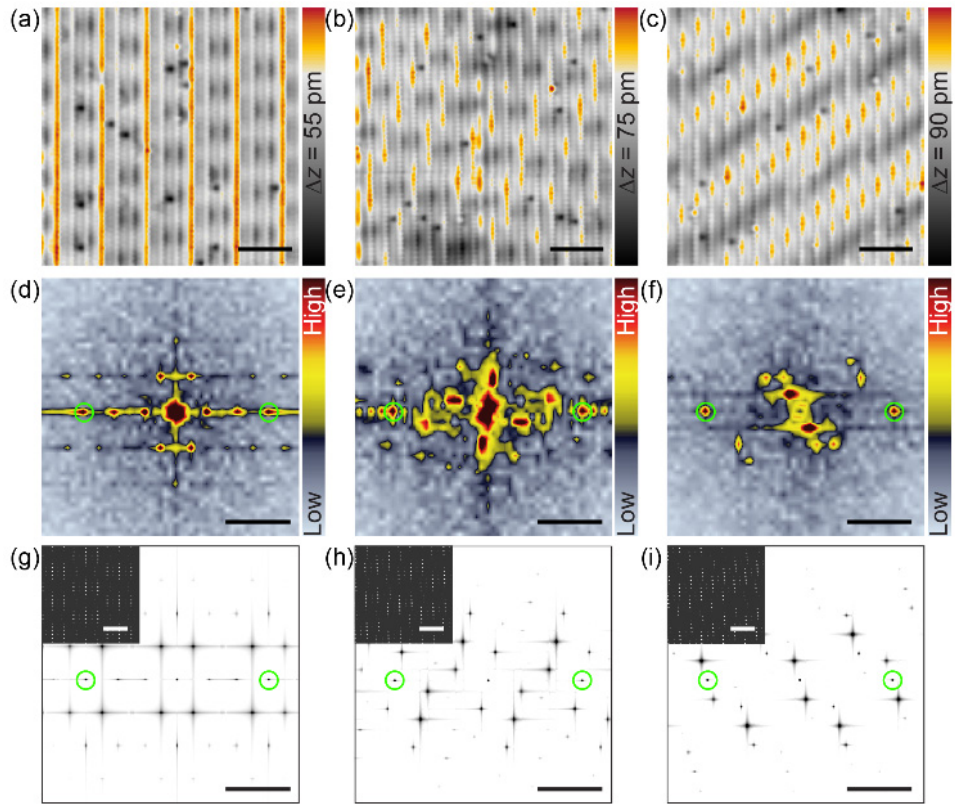

Figure 2. (a-c) Atomically resolved constant-current STM images showing identical atomic lattices with different moiré structures resulting from different rotation angles between the $V_2S_3$ layer and Au(111) substrate ($\theta = 0°$, $1.3°$, and $4.2°$ with respect to <110>) ($V_s = 1$ V, $I_t = 40$ nA, $T = 1.3$ K, scale bar = 3 nm). (d-f) FFTs of the respective images in (a-c) (scale bar = 5 nm$^{-1}$). (g-i) FFTs of simulated moiré images with the same scale bar. Insets show the simulations in real space (scale bar = 2 nm). Green circles highlight reciprocal lattice points of $V_2S_3$.

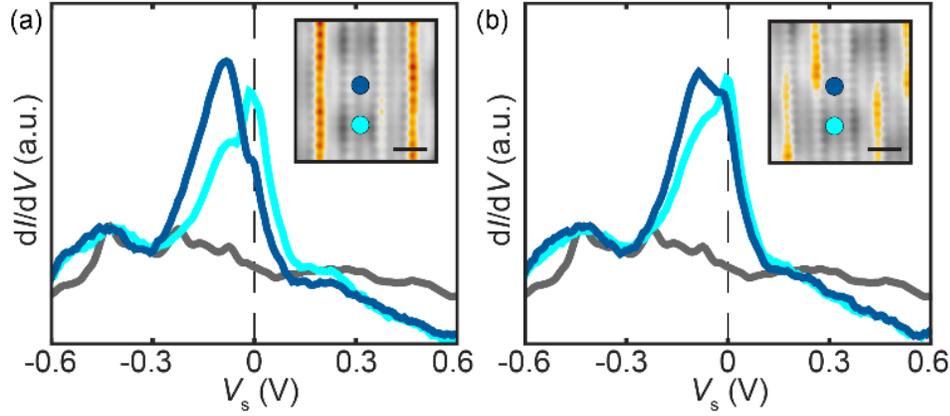

Figure 3. d$I$/d$V$ spectra acquired at different positions (indicated in blue in the insets) of the moiré superstructure of monolayer V$_2$S$_3$ on Au(111), for (a) θ = 0° and (b) θ = 1.3°. The peak in the d$I$/d$V$ is observed at all probed positions of the moiré unit cell, but with variations in energy. The surface state of Au(111) is shown for reference (gray). (Stabilization bias ($V_{stab}$) = 1.5 V, stabilization current ($I_{stab}$) = 600 pA, modulation bias ($V_{mod}$) = 1 mV, $T$ = 1.3 K; insets: $V_s$ = 1 V, $I_t$ = 40 nA, $T$ = 1.3 K, scale bar = 1 nm).

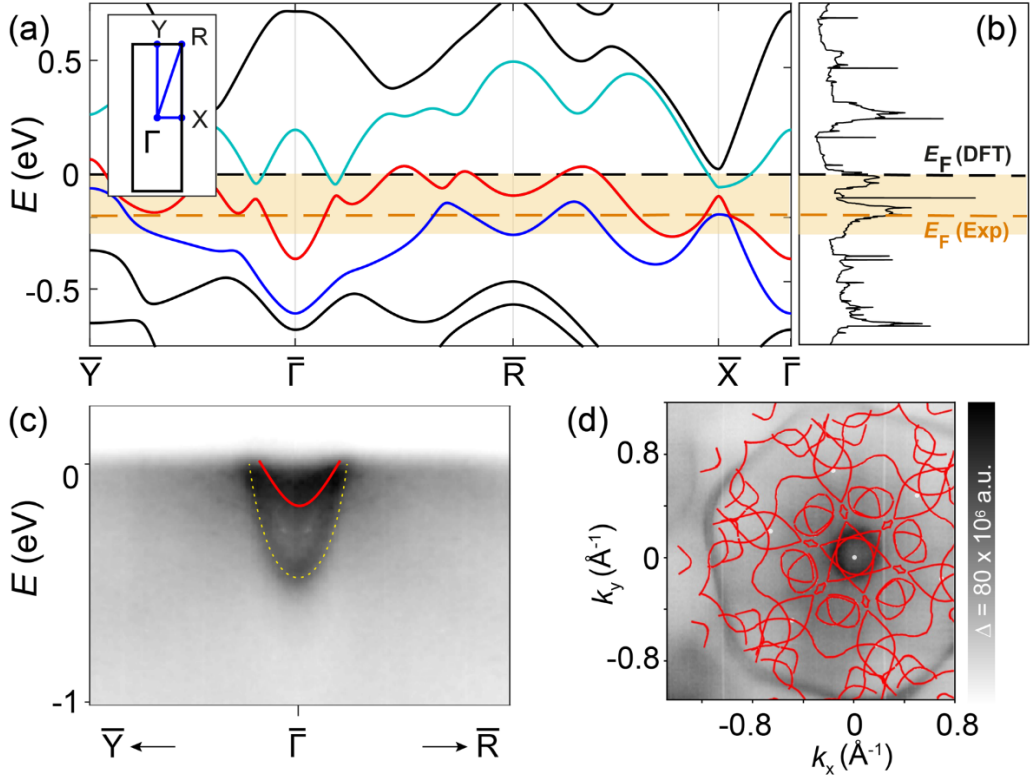

Figure 4. (a) Band structure of $V_2S_3$ along the high-symmetry path shown in the inset. The black dashed line indicates the DFT Fermi energy, while the Fermi energy determined by comparison with the ARPES measurements is marked by an orange dashed line. (b) Total density of states (DOS). (c) Comparison of the band dispersion around $\bar{\Gamma}$ measured by ARPES with DFT. Yellow dashed line illustrates the surface state of Au(111). (d) Fermi contour calculated from DFT taking into account the three rotational domains, in comparison with the ARPES measurements.

**Supporting Information**

**Moiré-induced electronic structure modifications in monolayer V₂S₃ on Au(111)**

Umut Kamber[1], Sahar Pakdel[2], Raluca-Maria Stan[2], Anand Kamlapure[1], Brian Kiraly[1], Fabian Arnold[2],

Andreas Eich[1], Arlette S. Ngankeu[2], Marco Bianchi[2], Jill A. Miwa[2], Charlotte E. Sanders[3], Nicola Lanata[2],

Philip Hofmann[2], Alexander A. Khajetoorians[1*]

1.    *Institute for Molecules and Materials, Radboud University, Nijmegen 6525AJ, The Netherlands*

2.    *Department of Physics and Astronomy, Interdisciplinary Nanoscience Center, Aarhus University,*
      *8000 Aarhus C, Denmark*

3.    *Central Laser Facility, STFC Rutherford Appleton Laboratory, Harwell, Didcot OX11 0QX, United*
      *Kingdom*

## 1. Moiré simulations

We simulated the expected moiré patterns in MATLAB by multiplying sinusoidal functions representing each of the atomic lattices. The Au(111) surface in Figure S1a is generated using three sinusoidal functions with the same wavelength, rotated 120º with respect to each other, as explained in Zeller *et.al* [1]. Similarly, top sulfur layer of $V_2S_3$ is generated using the function:

$$f_S(x, y) = \cos(2k_b x) + \cos(k_b x) + cos(k_a y)$$

where $k_a = 2\pi/a$ and $k_b = 2\pi/b$. The corrugation of both $f_S$ and $f_{Au}$ functions were chosen to be between 0 and 1. Both lattices were simulated on a 600nm × 600nm grid, with 2048 pixels in each direction. The real space simulations of both individual lattices, and their corresponding FFTs are shown in Figure S1. The resulting moiré pattern was calculated as a product of two lattices with orders *n* and *m*;

$$M = f_S^n \times f_{Au}^m$$

For the moiré superstructure shown in Figure S1c, and in all the simulations shown in the main text, 6th order functions are used, in order to capture all possible higher-order modes.

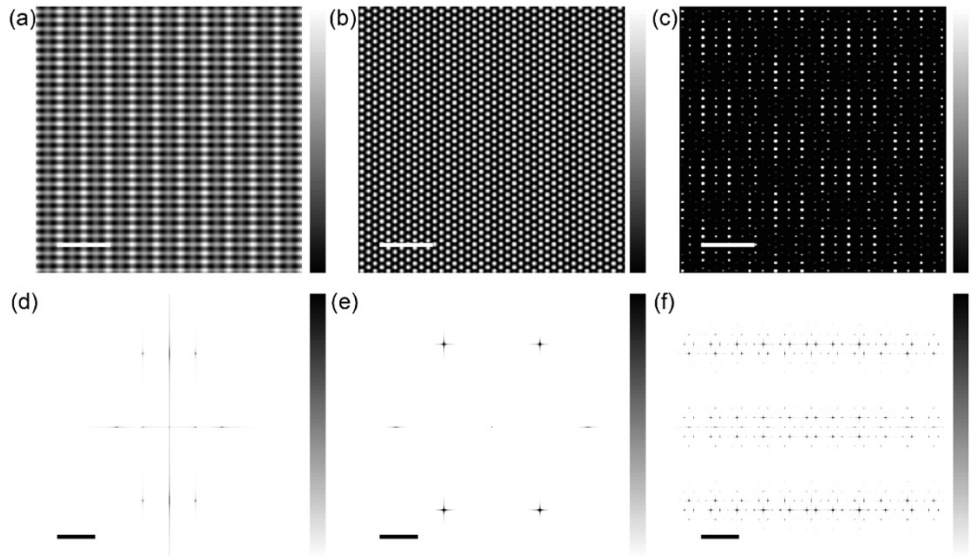

**Figure S1.** Simulations in real space for (a) top sulfur layer of $V_2S_3$ monolayer, (b) Au(111) surface, and (c) resulting moiré pattern ($n = m = 6$, scale bar = 2 nm). (d-f) Corresponding FFTs of the images above (scale bar = 10 nm$^{-1}$).

## 2. Real space characterization

A high-resolution STM image in Figure S2a shows the morphology of the sample surface. Elongated $V_2S_3$ islands with moiré superstructure, as well as the herringbone reconstruction of the clean Au(111) surface, can be clearly seen. The figure illustrates that a majority of the islands are oriented to yield the $\theta$ = 0° moiré superstructure (where the **a** direction of $V_2S_3$ and <110> direction of the substrate are parallel). Furthermore, these domains form with three different rotations, consistent with the three-fold symmetry of the Au(111).

Atomically resolved STM images at two different biases are shown in Figure S2b,c. A distinct feature with a longer-range modulation can be seen in the real space STM images in Figure S2b: the S rows between moiré minima (red-yellow rows) have higher apparent height than their neighbors. Following the moiré periodicity, these rows repeat every three-unit cells along the **b** direction, thus, they correspond to the A-point in reciprocal space.

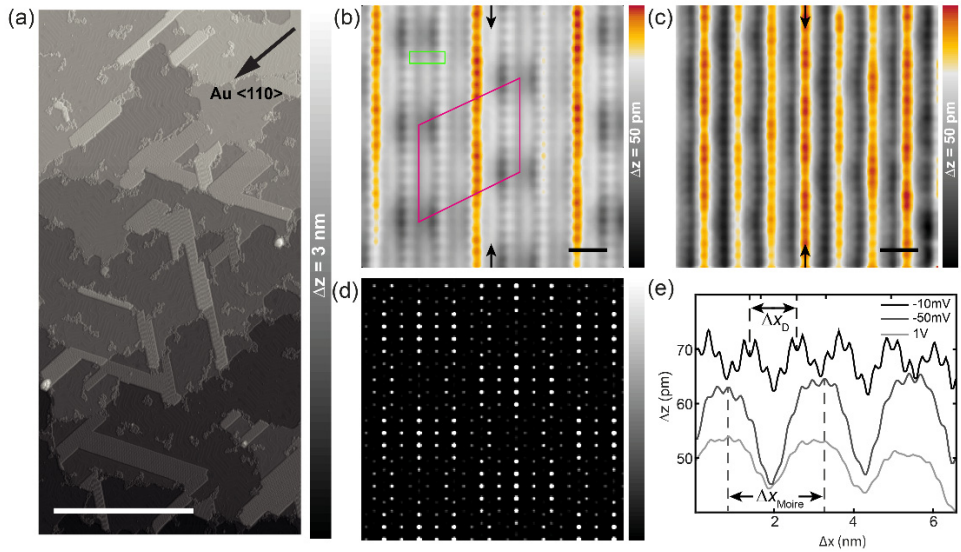

**Figure S2.** (a) Large scale STM image of the surface ($V_s$ = -1 V, $I_t$ = 100 pA, $T$ = 1.3 K, scale bar = 100 nm). (b) Atomically resolved STM image of the $V_2S_3$ island surface measured at $V_s$ = 1 V ($I_t$ = 40 nA, $T$ = 1.3 K, scale bar = 1 nm). (c) The exact same area is measured at $V_s$ = -10 mV. (d) Real space moiré simulations for the moiré superstructure in (b). (e) Line profiles extracted along the sulfur row (marked with black arrows in (b, c)) from the images measured at different biases.

The exact same area is measured at $V_s$= -10 mV (Figure S2c), showing that the contrast of the moiré superstructure disappears at low biases, while another longer-range modulation appears. Line profiles along the sulfur row were extracted from the images in Figure S2b,c and shown in Figure S2e, illustrating the apparent height modulations of individual periodicities together with atomic corrugation. The additional modulation observed at low bias imaging has a periodicity matching the one of the higher-order modes of the moiré superstructure (indicated as D-point in Figure 1). Real space simulations for the same moiré superstructure are shown in Figure S2d.

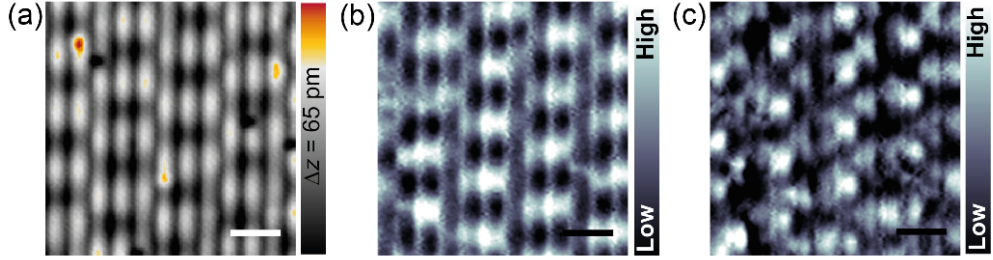

**Figure S3.** (a) Constant-current STM image ($V_s$ = -0.8 V, $I_t$ = 200 pA, $T$ = 1.3 K, scale bar = 2 nm). Spatially dependent differential conductance maps of the same area measured at (b) $V_s$ = -50 mV and (c) $V_s$ = -10 mV ($V_{mod}$ = 2 mV, $I_t$ = 200 pA, $T$ = 1.3 K, scale bar = 2 nm).

In order to capture spatial variations in the LDOS, we acquired differential conductance maps using a lock-in detection technique (Figure S3) and tunneling spectra on along the direction of atomic sulfur rows across multiple moiré units (Figure S4). A constant-current image of the surface is shown in Figure S3a, illustrating the atomic top sulfur rows as well as the moiré superstructure. Differential conductance maps were acquired in the same area at two different biases ($V_s$ = -50 mV in Figure S3b and $V_s$ = -10 mV in Figure S3c), where characteristic changes were observed in tunneling spectra shown in the main text (Figure 3 and Figure S4). As a result of the variations in LDOS within the moiré unit, seen in the shifting d$I$/d$V$ peak at approximately -100 mV in Figure S4, the differential conductance maps show a complicated real-space structure (Figure S3).

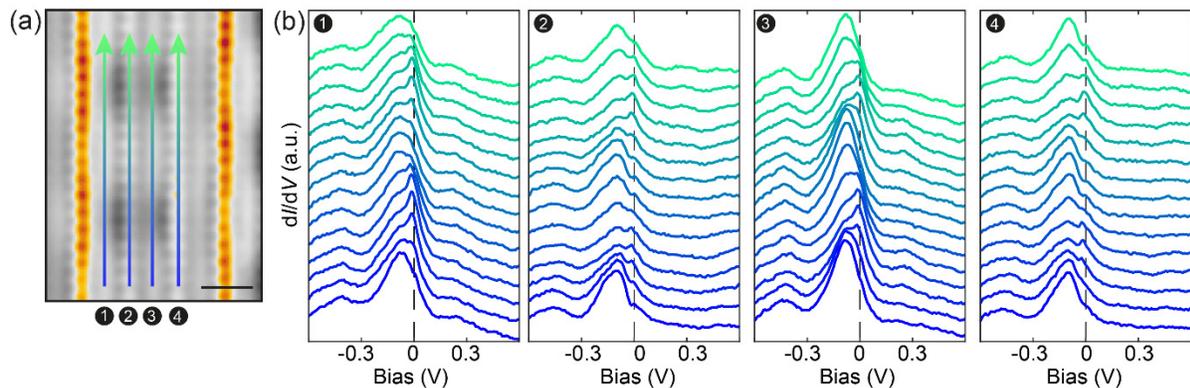

**Figure S4.** (a) Constant-current STM image ($V_s$ = 1 V, $I_t$ = 40 nA, $T$ = 1.3 K, scale bar = 1 nm). (b) d$I$/d$V$ spectra acquired at different positions along atomic sulfur rows (indicated by arrows and numbers) on the surface ($V_{stab}$ = 1.5 V, $I_{stab}$ = 600 pA, $V_{mod}$ = 1 mV, $T$ = 1.3 K).

### 3. Orbital decomposition

Projections of the total DOS onto each of the $d$-orbitals of vanadium is shown in Figure S5. The figure reveals that the character of the DOS near the experimental Fermi energy has a strong contribution from the out-of-plane V $d$-orbitals.

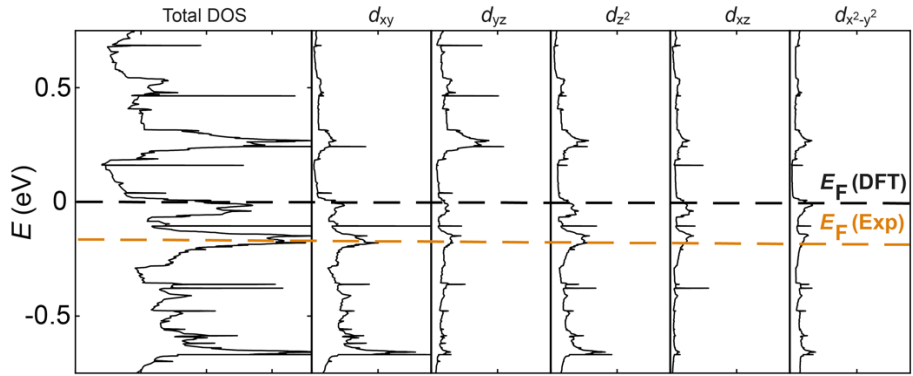

**Figure S5.** Total DOS and orbitally resolved contributions from the $d$-orbitals of vanadium.

### 4. Comparison between ARPES and DFT calculations

Here we discuss the behavior of the DFT and ARPES energy contours at different energies. In particular, in Figure S6, we compare the calculated DFT energy contours to the ARPES energy contours at two different energies, namely at the Fermi level $E_F$(exp) and at $E_F$(exp)−125 meV. Note that, besides effectively shifting the Fermi level, the V$_2$S$_3$ monolayer possibly hybridizes with the Au(111) substrate. Such hybridization effects are not included in these DFT calculations. Furthermore, since the ARPES intensity is highly dependent on matrix elements, not all features predicted by theory are visible experimentally. However, as explained below, some of the main qualitative features of the constant energy contours are consistent with the theoretical predictions, both at the Fermi level and at $E_F$(exp) −125 meV.

In Figure S6, we inspect the behavior of the band structure as a function of the energy, focusing on the V-$3d$ pockets around the $\bar{\Gamma}$ point and its second-order repetitions. Panels (c) and (g) show the DFT contributions to the constant energy contours originating from a single rotational domain (the one corresponding to the yellow rectangle in panel i), while panels (d) and (h) show the contributions from all rotational domains, simultaneously. At the first $\bar{\Gamma}$ point, the Au(111) surface state dominates the ARPES intensity in the constant energy contours. However, the behavior of the ARPES V-3$d$ pockets can be observed very clearly at the second-order Brillouin-zone repetitions of the $V_2S_3$ $\bar{\Gamma}$ point (marked by white dots in panels (b-d, f-h) of Figure S6), where the Au(111) background intensity is relatively weak. By comparing the panels (b) and (f), it is clear that the ARPES signatures narrow down around the indicated white dots when going from the $E_F$(exp) to $E_F$(exp) $-125$ meV. Based on the comparison in the panels (d) and (h), we attribute this behavior to the fact that the V-3$d$ pockets around the $\bar{\Gamma}$ point become smaller as the absolute energy is reduced.

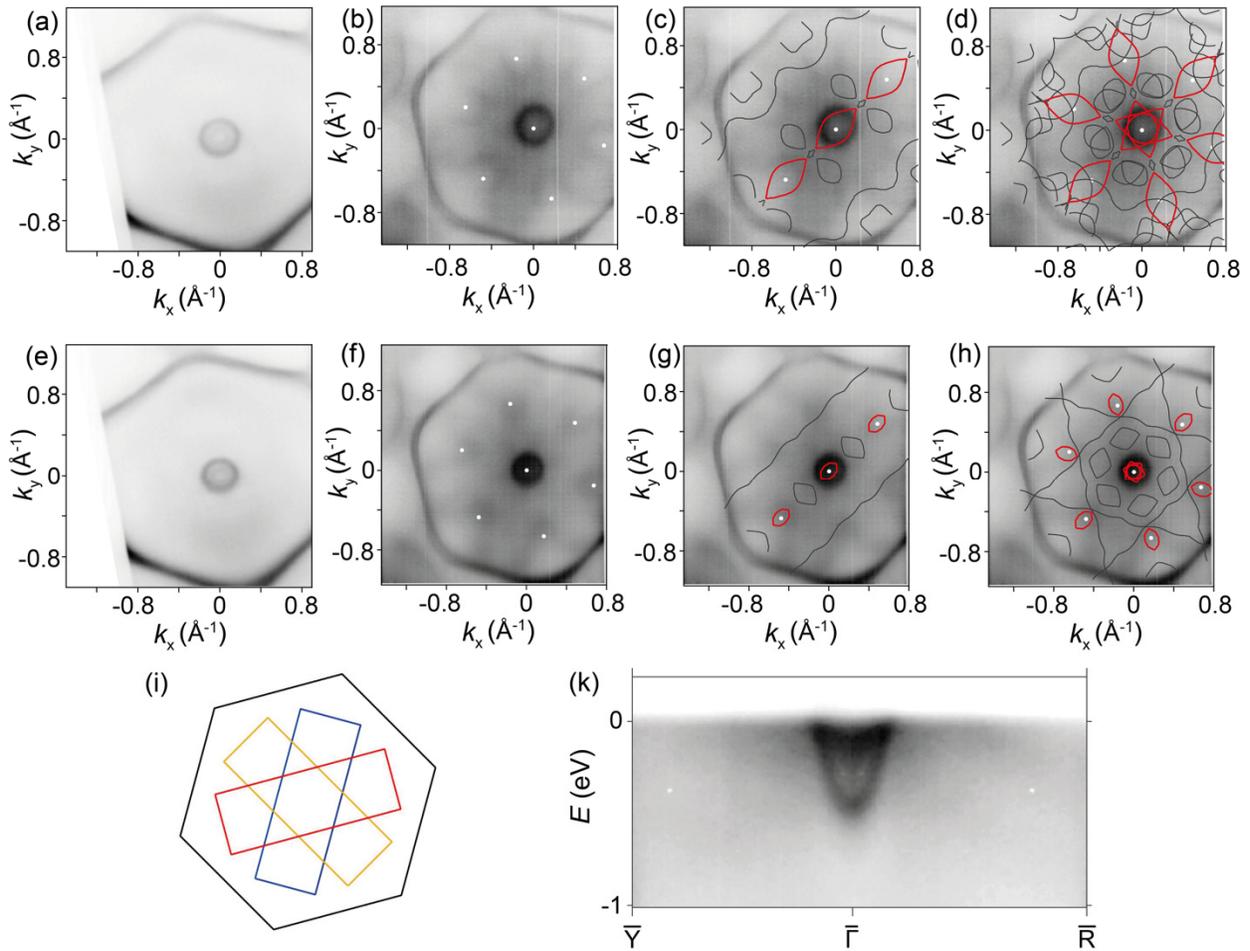

**Figure S6.** ARPES Fermi contour of (a) bare Au(111), and (b) $V_2S_3$/Au(111) at the Fermi level. (c-d) Fermi contour of the free-standing monolayer $V_2S_3$ calculated with DFT at $E_F$ (Exp), in comparison with ARPES data from (b). ARPES constant energy contours of (e) bare Au(111), and (f) $V_2S_3$/Au(111) at -125 meV. (g-h) Constant energy contours of the free-standing monolayer $V_2S_3$ calculated with DFT at $E_F$ (Exp) = -125 meV, in comparison with ARPES. In panels (c, g) only the contribution of one of the rotational domains is shown (yellow rectangle in panel i). (i) Schematics of the surface BZ of Au(111) and the BZs of the three rotational domains of $V_2S_3$ monolayer, with orientation consistent with panels. (k) Band dispersion

measured by ARPES on $V_2S_3$/Au(111), taken from Fig. 4. The white dots are referenced in the discussion above.

## 5. Electronic susceptibility calculations

To investigate the interplay between electronic and orbital degrees of freedom, we calculated the electronic susceptibility $\chi(\boldsymbol{q})$, as peaks in its real part $\chi'(\boldsymbol{q})$ and imaginary part $\chi''(\boldsymbol{q})$ (nesting function) are often associated with structural instabilities. Within the so-called "constant matrix element approximation"[2, 3], $\chi'(\boldsymbol{q})$ is given by;

$$\chi'(\boldsymbol{q}) = \sum_{\boldsymbol{k}} \frac{f(\epsilon_{\boldsymbol{k}}) - f(\epsilon_{\boldsymbol{k}+\boldsymbol{q}})}{\epsilon_{\boldsymbol{k}} - \epsilon_{\boldsymbol{k}+\boldsymbol{q}}}$$

(often referred to as the Lindhard function), where $\boldsymbol{k}$ belongs to the BZ, $\boldsymbol{q}$ is a generic reciprocal-space vector, $f$ is the Fermi function and $\epsilon_{\boldsymbol{k}}$ is the band energy.

The results of our calculations are summarized in Figure S7. Both the electronic susceptibility χ'(q) and the nesting function χ''(q), calculated at the theoretically estimated $E_F$ (DFT) for the free-standing $V_2S_3$ monolayer, display a peak very close to one of the second-order points of the $\theta$ = 0° moiré superstructure (indicated as the C-point in Figure S7 and in Figure 1). As these calculations only account for the free-standing monolayer, the modifications in the band structure induced by the moiré superstructure or the hybridization with the substrate are expected to yield considerable changes in the electronic susceptibility.

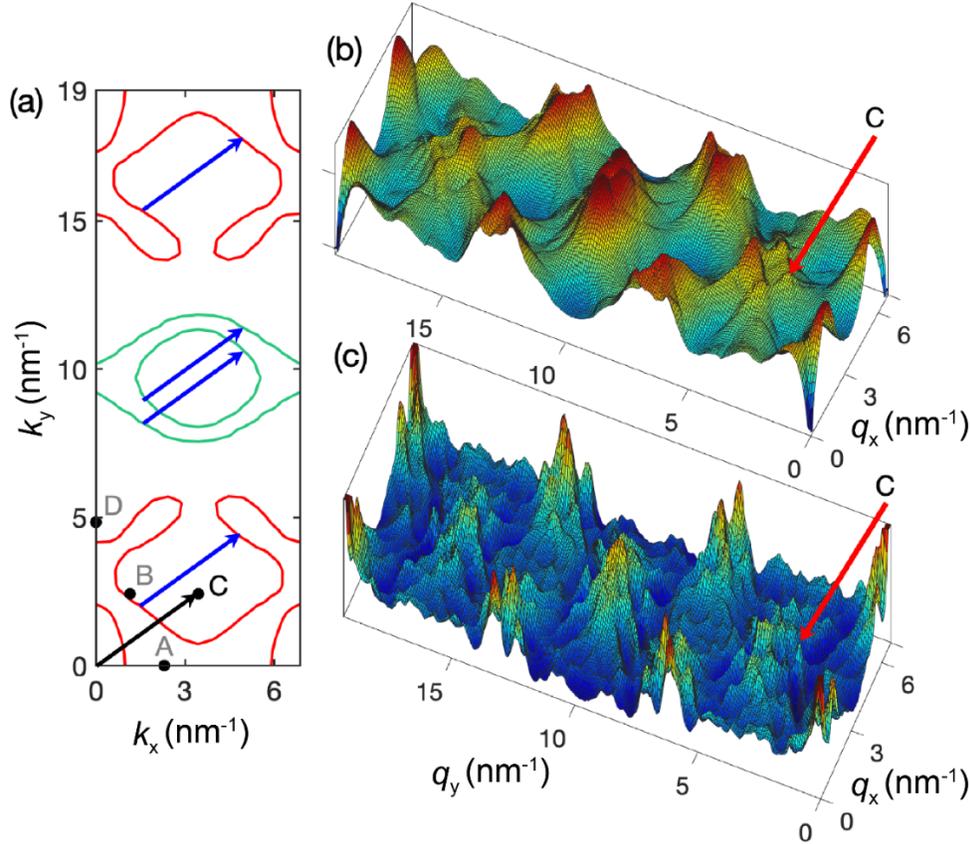

**Figure S7.** (a) Fermi contour of free-standing monolayer $V_2S_3$ calculated with DFT. The reference Fermi energy is set to the value $E_F$ (DFT) determined theoretically, see the black dashed line in Figure 4. The colors of the Fermi contour lines correspond to the bands shown in Figure 4. The black dots correspond to the reciprocal-space points indicated in Figure 1. (b) Real part of the electronic susceptibility $\chi'(\boldsymbol{q})$. (c) Imaginary part of the electronic susceptibility (nesting function) $\chi''(\boldsymbol{q})$. The Fermi-surface nesting (indicated by arrows in panel (a)) responsible for the peaks in $\chi(\boldsymbol{q})$ (panels (b, c)) coincides with the reciprocal higher order moiré point C.

## Supplementary References